\documentclass[twocolumn,aps,prl,showpacs,superscriptaddress,floatfix]{revtex4-1}
\usepackage[dvips]{color} 
\usepackage{graphicx} 
\usepackage{bm}

\begin{document}

\title{Heavy-Particle Radioactivities of Superheavy Nuclei}

\author{D. N. Poenaru}
\email[]{poenaru@fias.uni-frankfurt.de}
\author{R. A. Gherghescu}
\affiliation{Frankfurt Institute for Advanced Studies (FIAS),
Ruth-Moufang-Str. 1, 60438 Frankfurt am Main, Germany}
\affiliation{
Horia Hulubei National Institute of Physics and Nuclear
Engineering (IFIN-HH), \\P.O. Box MG-6, RO-077125 Bucharest-Magurele, Romania}
\author{W. Greiner}
\affiliation{Frankfurt Institute for Advanced Studies (FIAS),
Ruth-Moufang-Str. 1, 60438 Frankfurt am Main, Germany}

\date{ }

\begin{abstract}
The concept of heavy-particle radioactivity (HPR) is changed to allow
emitted particles with $Z_e>28$ from parents with $Z>110$ and daughter
around $^{208}$Pb. Calculations for superheavy (SH) nuclei with Z=104-124
are showing a trend toward shorter half-lives and larger branching ratio
relative to $\alpha$~decay for heavier SHs. It is possible to find regions
in which HPR is stronger than alpha decay. The new mass table AME11 and the
theoretical KTUY05 and FRDM95 masses are used to determine the released
energy. For 124 we found isotopes with half-lives in the range of ns to ps.
\end{abstract}


\maketitle

In recent years the heaviest elements with atomic numbers up to
$Z=118$ have been synthesised \cite{hof10p} either with cold fusion
reactions having the $^{208}$Pb or $^{209}$Bi target
\cite{hof00rmp,mor04jpsj} or with hot fusion induced by $^{48}$Ca
projectiles \cite{oga07jpg,oga10prl}.  Attempts to produce $Z=120$ are
reported \cite{oga09pr} and new experiments are presently running at GSI
Darmstadt \cite{hof11pc}.  The main experimental difficulty in identifying
the new superheavy (SH) elements is the low probability of their formation,
and the separation of the short lived compound nucleus from the very
high flux of incident projectile nuclei.  The lowest cross-section of 55~fb
was measured at RIKEN \cite{mor04jpsj} where one decay chain of $^{278}$113
was observed during 79 days with beam of $^{70}$Zn on $^{209}$Bi target. 
After naming copernicium, Cn, $Z=112$ suggested by GSI scientists, IUPAC
recommends that the Dubna-Livermore collaboration be credited with discovery
of new elements 114 and 116.

It is generally agreed that the term SH element, introduced
\cite{wer58pr} in 1958, is a synonym for elements which exist solely due to
their nuclear shell effects.  The lightest SH is $Z=104$ Rf with half-lives
of different isotopes around 1 min.  This is 16 orders of magnitude
longer than the expected nuclear lifetime of $10^{-14}$~s these isotopes
would survive without any shell stabilisation.
Spontaneous fission, the dominating decay mode in the region around Rf,
becomes a relatively weaker branch compared to $\alpha$-decay for the
majority of recently discovered proton-rich nuclides.  Extensive
calculations of fission barriers and half-lives have been published
\cite{mol09prc}.

Despite the important experimental and theoretical development there are
still several unanswered questions related to the magic numbers, production
cross sections, and decay modes.  Besides beta decay, only alpha decay and
spontaneous fission of SH nuclei have been experimentally observed up to
now.  We would like to take also into account heavy-particle radioactivities
(HPR) \cite{ps84sjpn80,enc95}.

Since 1984 \cite{ros84n}, the following HPR have been experimentally
confirmed \cite{bon07rrp} in heavy parent nuclei with $Z=87-96$: $^{14}$C,
$^{20}$O, $^{23}$F, $^{22,24-26}$Ne, $^{28,30}$Mg, $^{32,34}$Si with
half-lives in good agreement with predicted values within analytical
superasymmetric fission (ASAF) model (see the review \cite{p302bb10} and
references therein).  Almost always the corresponding daughter nucleus was
the doubly magic $^{208}_{82}$Pb$_{126}$ or one of its neighbors.  The
newest measurement of $^{14}$C decay of $^{223}$Ac \cite{gug08jpcs} was one
of the possible candidates for future experiments mentioned in the
systematics \cite{p240pr02} showing that the strong shell effect due to
magic number of neutrons, $N_d=126$, and protons, $Z_d=82$, present in order
to lead to shorter half-lives was not entirely exploited.

The shortest half-life of $T_c=10^{11.01}$~s corresponds to $^{14}$C
radioactivity of $^{222}$Ra and the largest branching ratio relative to
alpha decay $b_\alpha =T_\alpha/T_c=10^{-8.9}$ was measured for the $^{14}$C
radioactivity of $^{223}$Ra.  Consequently HPR in the region of transfrancium
nuclei is a rare phenomenon in a strong background of
$\alpha$ particles.  Several attempts to detect $^{12}$C radioactivity of
the neutrondeficient $^{114}$Ba with a daughter in the neighbourhood of the
double magic $^{100}_{50}$Sn$_{50}$, predicted to have a larger $b_\alpha$,
have failed.

In order to check the possibility of extrapolations from $A_e=14-34$ emitted
clusters already measured in the region of emitters with $Z=87-96$ to SHs
up to 124, where one may find an emitted particle as heavy as $^{114}$Mo, we
estimated within ASAF model the half-life for $^{128}$Sn emission from
$^{256}$Fm ($Q=252.129$~MeV) and for $^{130}$Te emission from $^{262}$Rf
($Q=274.926$~MeV): $\log _{10} T_{Fm}(s)=4.88$ and $\log _{10}
T_{Rf}(s)=0.53$.  They are in agreement with experimental values for
spontaneous fission \cite{hofd96mb}: 4.02 and 0.32, respectively.

There are many other theoretical approaches of the HPR e.g.  Refs. 
\cite{ble88prl,p195b96,lov98pr,qic09prl}.  Any spontaneous emission of a
charged particle from atomic nucleus may be explained as a quantum
mechanical tunnelling of a preformed cluster at the nuclear surface through
the potential barrier \cite{gam28zp}.  Microscopic calculations of cluster
formation probability and of barrier penetrability have been performed
\cite{lov98pr,qic09prl} by using the R-matrix description of the process. 
The half-life, $T_c$, is expressed as
\begin{equation}
T_c=\frac{\hbar \ln 2}{\Gamma_c}
\end{equation}
where $\Gamma_c$ is the decay width and h is the Plank constant. A universal
decay law for $\alpha$~emission and HPR was recently developed \cite{qic09prl}
based on this theory.

We should change  the concept of HPR, previously \cite{p160adnd91}
associated to a maximum $Z_e^{max}=28$, allowing to preserve its main
characteristics in the regions of SH with $Z>110$ i.e.  in a systematic
search for HPR we shall consider not only the emitted particles with atomic
numbers $2<Z_e<29$, as in previous calculations, but also heavier ones up to
$Z_e^{max}=Z-82$, allowing to get for $Z>110$ an atomic number of the most
probable emitted HP $Z_e > 28$ and a doubly magic daughter around
$^{208}$Pb.

Calculations are performed within the ASAF model, very useful for the high
number of combinations parent-emitted cluster (more than $10^5$) in order
to check the metastability of more than 2000 parent nuclides with measured
masses against many possible decay modes.  We started with Myers-Swiatecki
liquid drop model (LDM) \cite{mye66np} adjusted with a phenomenological
correction accounting for the known overestimation of the barrier height and
for the shell and pairing effects in the spirit of Strutinsky method.

The half-life of a parent nucleus $AZ$ against the split into a HP or an
emitted cluster $A_eZ_e$ and a daughter $A_d Z_d$ is given by
\begin{equation}
T = [(h \ln 2)/(2E_{v})] exp(K_{ov} + K_{s})
\end{equation}
and is calculated by using the Wentzel--Kramers--Brillouin (WKB) 
quasiclassical approximation, according to
which the action integral is expressed as
\begin{equation}
K =\frac{2}{\hbar}\int_{R_a}^{R_b}\sqrt{2B(R)E(R)} dR
\end{equation}
with $B=\mu$ the reduced mass, $K=K_{ov}+K_s$, and the $E(R)$ potential 
energy replaced by $[E(R)-E_{corr}] - Q$.
$E_{corr}$ is a correction energy similar to the Strutinsky \cite{str67np} 
shell correction, also taking into account the fact that LDM 
overestimates fission barrier heights, and the effective inertia in the 
overlapping region is different
from the reduced mass. $R_a$ and $R_b$ are the turning points of the WKB
integral.
The two terms of the action integral $K$, corresponding to the overlapping
($K_{ov}$) and separated ($K_s$) fragments, are calculated by analytical  
formulas \cite{p302bb10}.

Half-life calculations are very sensitive to the released energy (Q value)
obtained as a difference of the parent and the two decay product masses
\begin{equation}
Q=M - (M_e + M_d)
\end{equation} 
in units of energy.
Even with the newly released tables of experimental masses, atomic mass
evaluation 2011 (AME11)
\cite{aud11pc} as a preview for the AME13 publication, many masses are still
not available for new SH, hence we shall use not only these updated tables
for 3290 nuclides (2377 measured and 913 from the systematics) ending up at
$Z=118$ but also some calculated masses, e.g., Koura-Tachibana-Uno-Yamada
(KTUY05) \cite{kou05ptp} and the finite-range droplet model
(FRDM95) \cite{mol95adnd} with 9441 and 8979 masses, respectively.

\begin{figure}[htb] 
\includegraphics[width=7.6cm]{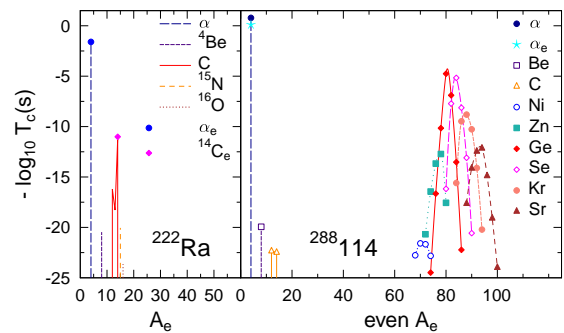} 
\caption{(Color online) Time spectra of different cluster emissions from 
$^{222}$Ra (left panel) and from the superheavy nucleus $^{288}$114 (right
panel). The most probable emitted clusters from $^{222}$Ra and $^{288}$114
are $^{14}$C and $^{80}$Ge, respectively, both leading to $^{208}$Pb
daughter nucleus.
\label{tspec}} 
\end{figure}

In a systematic search for HPR we calculate with the ASAF model for
every parent nucleus $^A$Z the half-lives of all combinations of pairs of
fragments $^{A_e}Z_e$, $^{A_d}Z_d$ with $2 < Z_e \leq Z_e^{max}$ conserving
the hadron numbers $Z_e+Z_d=Z$ and $A_e+A_d=A$.  
Let us start with the results obtained by using the AME11 mass tables.
An example of the time
spectra obtained for different clusters emitted from the parent nuclei
$^{222}$Ra and $^{288}$114 is shown in Fig.~\ref{tspec} versus the mass
numbers of the light fragment.  The symbols of the emitted HPR are given on
the figure's legend.  
\begin{figure*}[htb] 
\includegraphics[width=9.6cm]{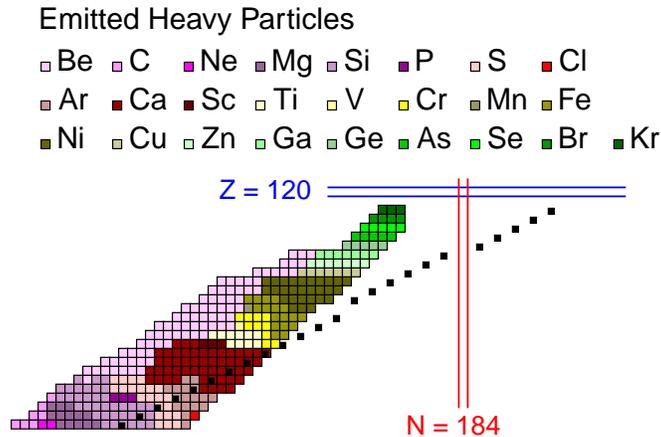} 
\caption{(Color online) Chart of heavy and superheavy cluster emitters with
atomic numbers $Z=94-118$.  The $Q$~values are calculated using the AME11
mass tables \cite{aud11pc}.
Black squares mark the Green approximation of the line of beta stability. 
One most probable emitted cluster is given for every parent nucleus.
\label{cha-ame11}} 
\end{figure*}

From the left panel of this figure one can see that
the shortest half-lives of $^{222}$Ra correspond to $\alpha$~decay and
$^{14}_6$C$_8$ radioactivity, respectively.  Both these decay modes have been
experimentally observed and there is a good agreement between the calculated
values and measured data.  Other HPR with half-lives $T_c < 10^{25}$~s are:
$^8$Be; $^{12,13}$C; $^{15}$N and $^{16}$O --- all with much longer
half-lives. 

Similarly, on the right hand side of the figure we show calculated results
for the SH nucleus $^{288}$114.  Again $\alpha$~decay is the strongest decay
mode and there is a good agreement between our calculations and the
experimentally observed half-life $T_\alpha$.  The time spectrum in the
region of mass numbers of emitted particles around $A_e=80$ is more complex
looking similar to a fission fragment spectrum.  There are many HPR with
$Z_e=28-38$ having $T_c < 10^{25}$~s.  For the sake of clarity we only
plotted the results corresponding to even-even emitted HP which are leading
to shorter half-lives in the same way the $^{13}$C radioactivity of
$^{222}$Ra is less probable than both $^{14}$C and $^{12}$C spontaneous
emissions.  The most probable emitted HP from $^{288}$114 is
$^{80}_{32}$Ge$_{48}$ with a calculated branching ratio
$b_\alpha=10^{-5.01}$.  One should also take into account a competition of
$^{84}_{34}$Se$_{50}$ with a magic number of neutrons $N_e=50$ and a
branching ratio $b_\alpha=10^{-5.42}$.

We proceed in a similar way with all parent nuclei with 
$Z=94-118$ present
on the AME11 mass table.  The chart of cluster emitters from
Fig.~\ref{cha-ame11} is obtained by associating to each parent only the
most probable emitted cluster.  The black squares mark the Green
approximation of the line of beta stability.  All superheavy nuclei present
on the AME11 mass table are proton-rich nuclides with neutron numbers
smaller than $N_\beta$ on the line of beta stability. The experimentally
determined $^{28}_{12}$Mg radioactivity of $^{236}_{94}$Pu, $^{32}_{14}$Si
radioactivity of $^{238}_{94}$Pu, and $^{34}_{14}$Si radioactivity of
$^{242}_{96}$Cm are fairly well reproduced. 

New many types of HPR with $Z_e>28$ may be seen on this chart: Cu, Zn, Ga,
Ge, As, Se, Br and Kr.  We used only one color for a given $Z_e$ despite the
fact that as the result of the calculations we obtained several isotopes of
these elements, e.g.  $A_e=26, 28$ for Mg; $30, 32, 33, 34$ for Si; $36, 38,
40, 41, 42$ for S; $44, 46, 47, 48$ for Ar; $48, 49, 50, 51, 52$ for Ca;
$50, 51, 52$ for Sc; $53, 54, 55, 56$ for Ti; $57, 58, 59, 60, 61$ for Cr;
$60, 62, 63, 64, 66$ for Fe; $66, 68, 69, 70, 71, 72, 73$ for Ni; $69, 71,
72, 73, 74, 75$ for Cu; $72, 74, 76, 77$ for Zn; $75, 77, 78, 79$ for Ga;
$78, 80, 81$ for Ge, $81, 83$ for As; $82, 84, 85$ for Se; $85, 86$ for Br,
and $86, 87$ for Kr.  Only one mass value was obtained for the most probable
emitted particles Be, C, Ne, P, Cl, V, and Mn.

\begin{figure}[htb] 
\includegraphics[width=8.6cm]{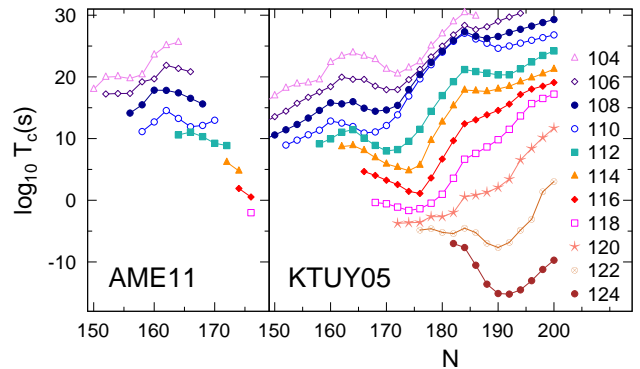} 
\caption{(Color online) Decimal logarithm of the half-lives of superheavy
nuclei against cluster radioactivities versus the neutron number of the
parent nucleus.  $Q$~values are calculated using the AME11 experimental
mass tables \cite{aud11pc} (left panel) and the KTUY05 \cite{kou05ptp} 
calculations.  
\label{tclus}} 
\end{figure}
\begin{figure}[htb] 
\includegraphics[width=8.6cm]{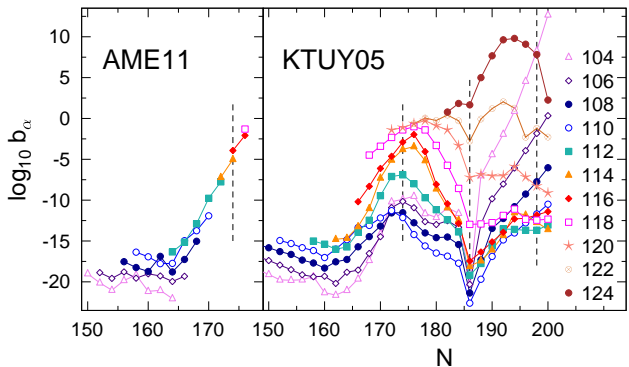} 
\caption{(Color online) Decimal logarithm of the branching ratio relative to
$\alpha$~decay for cluster emission from superheavy nuclei versus the
neutron number of the parent nucleus. Vertical dashed lines correspond to
$N=174, 186, 198$.
\label{bran}} 
\end{figure}

As we previously observed \cite{p160adnd91}, many of the proton-rich SH
nuclides are $^8$Be emitters, but they have a very low branching ratio
$b_\alpha$. The general trend of a shorter half-life and a larger branching
ratio when the atomic and mass numbers of the parent nucleus increases may
be seen on the left hand side of the figures~\ref{tclus} and \ref{bran},
obtained within ASAF model by using the AME11 mass tables to calculate the
$Q$~values. 

One can advance toward neutron-rich nuclei by using the KTUY05 calculated
mass tables, as shown in the right panels of these figures.  When using
KTUY05 and FRDM95 masses for parent and daughter nuclei 
we take into account the nuclides stable against
one proton, two protons, one neutron and two neutrons spontaneous emissions. 
If the calculated masses are reliable, then half-lives $T_c$ in the range of
nanoseconds to picoseconds for SH nuclei with $Z=124$ (see the right 
hand side of Fig.~\ref{tclus}) would make difficult or even impossible any
identification measurement.  More interesting for future experiments could
be some even-even proton-rich isotopes of the 122 element with $N=188-194$
for which the neutron number of the Green approximation of the line of beta
stability is $N_\beta=202$.

The pronounced minimum of the branching ratio at $N=186$ in
Fig.~\ref{bran} is the result of the strong shell effect of the assumed
magic number of neutrons $N=184$ present in the KTUY05 masses.  The
half-life of $\alpha $~decay of a SH nucleus with $N=186$ neutron number
leading to a more stable daughter with magic neutron number $N_d=184$ is
shorter by some orders of magnitude compared to the $\alpha $~decay of a SH
with $N=184$. Calculated branching ratios $b_\alpha > 1$ for Rf ($Z=104$)
only occur in very neutron-rich nuclei with $N=194-200$ compared to
$N_\beta=166$. Also their $T_c$ half-life is extremely long.
Similar results were obtained using the FRDM95 masses.

In conclusion, the concept of HPR should be changed to allow spontaneous
emission of heavy particles with atomic number larger than $28$ from SHs
with $Z>110$ and consequently daughter nuclei around the doubly magic
$^{208}$Pb.  The calculated half-lives $T_c$ against HPR and the branching
ratios relative to $\alpha$~decay $b_\alpha$ are showing a trend toward
shorter $T_c$ and larger $b_\alpha$ for heavier SH nuclei which are not
synthesised until now.  If the KTUY05 and FRDM95 masses used to calculate
the released energy $Q$ are reliable, we expect to find for the element 124
many isotopes with half-lives in the range of nanoseconds to picoseconds,
making practically impossible to perform any identification experiment. 
Nevertheless, there would be a chance to observe some proton-rich isotopes of
122 with branching ratios $b_\alpha > 1$.

We are looking forward to receive experimental information about the decay
modes of SHs with $Z > 120$, hoping to confirm the present calculations. 
There is also a need for developing more refined decay models as well as new
calculated mass tables and new mass measurements.

\begin{acknowledgments}
This work is partially supported by Deutsche Forschungsgemeinschaft Bonn and
partially within the IDEI and NUCLEU Programms under contracts with UEFISCSU
and ANCS, Bucharest.
\end{acknowledgments}


\end{document}